\documentclass[10pt]{article}
\usepackage[cp1250]{inputenc}
\usepackage{amsmath}
\usepackage{amssymb}
\title{Pbit and other list sorting algorithms}
\author{David S. P\l aneta\footnote{dplaneta@gmail.com}}
\usepackage{pst-plot}
\usepackage{pstricks}

\begin{document}
\maketitle
\begin{abstract}
Pbit, besides its simplicity, is definitely the fastest list sorting algorithm. It considerably surpasses all already known methods. Among many advantages, it is stable, linear and be made to run in place. I will compare Pbit with algorithm described by Donald E. Knuth~\cite{Knuth} in the third volume of ``The Art of Computer Programming'' and other list sorting algorithms.
\end{abstract}

\section{Introduction to lists}
A Lists is a set of independent data formats, very often called nodes. Particular nodes are connected with each other by means of pointers. Nodes are usually created dynamically, which is good for economical use of memory. We will deal with singly-linked lists because Pbit was created for such type of lists (which does not prevent from sorting double-linked lists). In a singly-linked list each node contains pointer only for another node in the list. The last pointer shows NULL. To remember at which point the list starts there is a special pointer called root or a head of the list~\cite{Ken}. The pointer of root may also be the first node of the list. Queue or Stack are examples of a singly-linked list. 

\section{Structure of data representing a node and description of a simply function inserting into the singly-linked list.}
\begin{center}
\begin{tabular}{|l|}
\hline
\textbf{struct} \verb@Node{@\\
\verb@Node *next;@\\
\verb@TYPE data;@\\  
\verb@};@\\
\hline
\end{tabular}
\parbox[c]{40ex}{
A node is the structure which contains at last one pointer for `node' and field of any `data' type, on the ground of which the nodes will be sorted.}
\end{center}
We know what a particular of the list looks like, so let us analyse the function inserting elements into it. Function `\verb@push@' corresponds with the function stacking elements.
\begin{center}
\begin{tabular}{|l|l|}
\hline	
1 & \textbf{void} \verb@push(Node **root, TYPE value){@\\
2 & \verb@Node *new_Node = @\textbf{new}\verb@(std::nothrow) Node;@\\
3 & \textbf{if}\verb@(new_Node == NULL){@\\
4 & \emph{//Error -- memory overflow.}\\
5 & \textbf{return;}\\
  & \verb@}@\\
6 & \verb@new_Node->data = value;@\\
7 & \verb@new_Node->next = *root;@\\
8 & \verb@*root = new_Node;@\\
  & \verb@}@\\
\hline	
\end{tabular}		
\end{center}
\begin{enumerate}
	\item Function `\verb@push@' takes two parameters. One is the pointer of the first element of the list (root) and the other argument is the value added to the list.
	\item We declare pointer of `\verb@node@' type, next we attribute to it memory declared by the `\textbf{new}' operator.
	\item If the pointer shows NULL it means there is no space in memory (memory overflow).
	\item Function should send the message about an error and \ldots 
	\item Finish working.
	\item Haring dealt with storage allocation, we write in the new node the value which was passed by function argument.
	\item Node `\verb@next@' is ascribed address of the root (head) that is attributed to first element of the list.
	\item The root indicate at the new node.	
\end{enumerate}

\section{The idea of sorting using Pbit}
The whole idea of Pbit is very interesting. Besides its simplicity it is indisputable the fastest algorithm which sorts lists. Among many advantages, it is stable, linear and does not require memory (``in situ'' sorting algorithm).\\
\rule{\textwidth}{0.1pt}
\textbf{Split the list in relation to the bit pattern. Each pattern variation is represented by separate list. Apply recurrently the same sorting function to all the lists but remember to project the pattern onto the succeeding bits. When the last key is looked through, merge \emph{sorted fragments} of the list.}
\\
\\
\emph{Pbit is combination of BucketSort~\cite{BucketSort}, RadixSort~\cite{RadixSort} and MergeSort~\cite{MergeSort} ideas.}
\\
\rule{\textwidth}{0.1pt}
\newline
Bit pattern is a set of \textbf{\emph{K}} bits.
\textbf{\emph{K}} defines the number of bits which are cut off the binary word.
\textbf{\emph{M}} defines number of bits in a binary word. For example, one-byte sign 
type 'char' contains $8$ bits. Four-byte number is $32$ bits long. 
$\Omega$ is variation \textbf{\emph{K}} of element binary set \{0,1\}, $\Omega = 2^K$. 
It determines the number of groups (number of roots), which the list may be divided into during one step. 
\textbf{\emph{n}} is the number of sorted elements. Meaningless for linear Pbit sorting.
Coefficient ``by \textbf{\emph{n}}'' depends only on 
\textbf{\emph{K}} (length of bit pattern) and \textbf{\emph{M}} (number of bits in binary word).
Coefficient ``by \textbf{\emph{n}}'' can be determined by dividing \textbf{\emph{M}} by \textbf{\emph{K}}.
If using my merging method, add \textbf{\emph{n}} to the coefficient.
\newline
$f(n) = \frac{M}{K} * n + n$\\
In Pbit I use four-bit pattern ($K=4$) for short lists, which means that I split the list into sixteen groups in 
maximum($\Omega = 2^K =  2^4  = 16$).
\newline
Constants in (source code) Pbit sorting algorithm:\\
\textbf{\emph{K}} = $4$ (four-bit pattern is excellent for short lists sorting)\\
$\Omega$ - dependent on \textbf{\emph{K}}; equals ($2^K = 16$)\\
Variables (depending on type of data):\\
\textbf{\emph{M}} - depends on the size of sorted element.\\
\textbf{\emph{n}} - depends on the number of elements.
\newline
Example:
We have a list consisting of eight one-byte elements: \\
$21 \longrightarrow 3\longrightarrow 209 \longrightarrow 14 \longrightarrow 156 \longrightarrow 47 \longrightarrow 3 \longrightarrow 214 \longrightarrow$ NULL
\newline
We put element of the list in relation to the first four bits ($K=4$) from
the right (the most important -- little-endian order of byte):
\newline
\{3,14,3\} | \{21\} | \{47\} | \{156\} | \{209,214\}
\newline
Next, we put elements in each groups separately looking at the following four bits 
(in case of one-byte type - the last ones):
\newline
\{ \{3\} | \{3\} | \{14\} \}   |   \{21\}   |   \{47\}   |   \{156\}   |   \{ \{209\} | \{214\} \}
\newline
In ``\emph{two steps}'' \textbf{\emph{n}} one-byte elements are sorted.
\newline
\newline
\emph{In how many steps four-byte numbers will be sorted?}
\newline
In $\frac{M}{K}$.
Number of bits of sorted element divided by the length of bit pattern.
The length K of the pattern is constant which equals 4. In case of four-byte elements
variable \textbf{\emph{K}} equals 32 ( because $4$ bytes = $4*8$ bits = $32$) bits.
So coefficient ``by \textbf{\emph{n}}'' equals $\frac{32}{4} + 1 = 8 + 1 = 9$. 
In eight steps four-byte numbers are sorted. In one step the lists are merged.
\newline
\newline
\emph{
O(n) algorithms such as countsort or radixsort required memory proportionally to \textbf{\emph{n}}. 
How is it possible that \emph{Pbit does not require memory proportionally to} \textbf{\emph{n}}?
}
Pbit does not count how many times number `x' occurred in sequence `L'. It does not compare numbers, (with one another) either. This algorithm is based on \textbf{a \emph{specific grouping} of elements} and needs only a few bytes parameters or roots. It is a very little amount of memory independent of  \textbf{\emph{n}} --- number of elements to be sorted. Precisely, it depends on $\Omega$ and coefficient.
\newline
The maximum amount of occupied storage can be expressed by formula:
\newline
$T = (\Omega*4 + 3*4) * \frac{M}{K}  + 2*4$ bytes, assume size of pointer equal four bytes
\newline
For constant \textbf{\emph{K}} which equals four, $\Omega$ equals sixteen.
\newline
$T = (64+12) * \frac{M}{4} + 8$ bytes
\newline
$T = 19 * M + 8$ bytes
\newline
The amount of occupied storage \textbf{\emph{T(M)}} will depend only on the size of sorted element.
\newline
For sign types (1 bytes = 8 bits) Pbit will require maximum $T(8) = 160$ bytes. 
\newline
For ``integers'' (4 bytes = 32 bits) Pbit will require maximum $T(32)= 616$ bytes.
\\
It means that it needs constant, modest space usage.\\
Space usage = $\Theta(\frac{M}{K}* \Omega)$\\
$\Omega = 2^K$, $K$ - constant\\
$M$ defines number of bits in a binary word (size of TYPE). There is $M$ constant.\\
That's why memory complexity equal O(1).\\
\textbf{The greatest advantages of Pbit is it be made to run in place.\\
It requires merely O(1) memory!}\\
\\
\emph{Why maximum instead of constant amount of memory?}
\newline
Because ``empty'' lists are omitted and are not taken into account during successive recurrent calls.
\newline
\newline
\emph{What order algorithm is responsible for merging list?}
\newline
It's O(n), because Pbit merges only those fragments of the list which were sorted. We group list elements till the last key. When the last key is looked through, all we have to do is to merge sorted fragments of the list. Every few, maximum $\frac{M}{K}$, recurrent calls sorted fragment of the list is (recursion) merged.
\newline
\newline
\emph{Is linearity the best advantage of Pbit?}
\newline
I don't think so. O(n) time algorithms rather will be much faster than O(nlgn) only while sorting very long lists.
\newline
The greatest advantage of Pbit does not require memory proportionally to \textbf{\emph{n}} (it's not extensive).
Pbit is stability and the fact that it can be adapted for sorting ``special data formats''.
In case of Pbit you can't talk about expected or pessimistic time complexity. Time complexity is always the some, no matter how many elements there are and what arrangement of sequence is (sorted, all elements are the some). 
In the four position I would put the statement about linear time complexity, small coefficient.
Last, but not last advantage of Pbit is simplicity in implementation and clear idea.
\newline

\section{Cutting bits off in binary words}
\begin{center}
\textbf{\emph{All quantities in computer are represented by binary words.}}
\end{center}
Pbit is based on cutting bits off and sorting elements in relation to those bits.
Unfortunately, not every high-standard language has built-in operators which carry out operations on bits. My means of arithmetic operators, however, it's possible to build a function which can cut bits off in a binary word.
Pbit will use function of cutting bits off in a binary word, which in Pascal language can be defined as follows:
\begin{center}
\begin{tabular}{@{} l @{}}
function bits(number, M: integer) : integer;\\
begin\\
\hspace{9 ex}bits := (number div ( $2^M$ ) ) mod  $\Omega$\\
end
\end{tabular}
\flushright{\textbf{\emph{M}} - bit number, from which the pattern in binary word is determined\break $\Omega = 16$}
\end{center}
Dividing by power of two is like shift bits to the right with index exponent.\\
For example, we divide $123 / 8 = 15,375$\\
$123$ decimally is $01111011$ binary.\\
We move bits three positions to the right (three because the third power of two equals eight): 
$000 01111$ binary is 15 decimally. In Turbo Pascal there is a special function for this purpose.\\
Language C has built-in operator ``\verb@>>@''.
\newline
\newline
\verb@number div@ $2^M$\\
in C can be written:\\
\verb@number >> M@\\
\newline
The result is divided by modulo $\Omega$. 
The operation of modulo division is carry in order to read only \textbf{\emph{K}} last bits ($\Omega = 2^K  =  16$).
The first versions of algorithm for cutting off bits in sorted elements used projection with bit field structure. To implement algorithm in other program languages, not only C/C++, I have chosen arithmetic/logic method.
\newline
\newline
\verb@bits := (number div (@$2^M$\verb@)) mod @$\Omega$\\
In C operations div and modulo can be changed into faster logical operations:
\verb@bits = (number >> M ) & (@$\Omega-1$\verb@)@, assume $\Omega = 2^K$
($\Omega-1$) - received number should be written hexadecimals.
\newline
\newline
I use bit shift, instead of division by power of two and logical operator `\verb@AND@', instead of modulo division because the statement which counts index of root table should be optimized to a maximum. It is this statement that the speed of function depends on.
\newline

\section{Implementation in C++}
\begin{center}
\begin{tabular}{|r|l|}
\hline	
 1& \verb@Node *Pbit(Node *L, @\textbf{unsigned}\verb@ M, Node *P=NULL){@\\
 2& \textbf{if}\verb@(L){@\\
 3& \hspace{7 ex}\verb@M=M-K;@\\
 4& \hspace{7 ex}\verb@Node *tab[@$\Omega$\verb@]={NULL};@\\
  & \\
 5& \hspace{7 ex}\textbf{for}\verb@(Node *i,**in; i=L; i->next=*in,*in=i){@\\
 6& \hspace{7 ex}\verb@in=&tab[(L->data>>M) % @$\Omega$\verb@];@\\
 7& \hspace{7 ex}\verb@L=L->next;@\\
  & \hspace{7 ex}\verb@}@\\
  & \\
 8& \hspace{7 ex}\textbf{if}\verb@(M) @\textbf{for}\verb@(@\textbf{int}\verb@ i=0; i<@$\Omega$\verb@; i++) P=Pbit(tab[i], M, P);@\\
 9& \hspace{7 ex}\textbf{else for}\verb@(@\textbf{int}\verb@ i=0; i<@$\Omega$\verb@; i++) P=merge(tab[i], P);@\\
  & \hspace{7 ex}\verb@}@\\
  & \\
10& \textbf{return}\verb@ P;@\\
  & \verb@}@\\
\hline	
\end{tabular}		
\\
Function takes parameters:\\
\parbox[t]{50ex}{
\textbf{L} -- the first node of the list to be sorted.\break
\textbf{P} -- ``end maker'', assumptive equals  NULL.\break
\textbf{M} -- size of sorted element expressed in bits.\break
}
\break
Function returns pointer to the first element of the sorted in decreasing order.\\
\end{center}

\begin{enumerate}
  \item \verb@Node *Pbit(Node *L, unsigned M, Node *P=NULL);@\\
  Function has three parameters, one of which assumptive equals NULL. Two parameters and function one of Node.
	Node is the structure which consists of at least pointer of `Node' type and variable `data'. 
  List will be sorted in relation to variable (key) `data'. 
  There is a broader description of lists and structure in the first and second point
	\item After calling the function we check whether the list which Pbit is to sort is not empty.
	Such check-up is not necessary but it makes the algorithm faster, because empty lists are omitted in the process of grouping. 	If the list happens to be empty we return pointer P (end marker), thus we return part of a sorted list. Additional, we could 		check if we do not sort one-element list because in case of one-element list it's not necessary to look through all bits of 		sorted element. However, it does not bring any significant benefits.
	\item Operation of subtraction of pattern length from length of sorted element is connected with subsequent grouping of 				nodes. At the first stage of sorting for four-byte numbers $M=32$ and constant $K=4$. $M=M-K$, $M=28$ -- which means that we 		will read determine list number into with $28$ (to 32) bit. Four bits which were read determine list number into which we 			insert sorted element.
	\item We create root table of new lists. Sorted nodes will be placed in those lists. Each elements of table is declared with 		NULL. \verb@table[size]={0,@\ldots\verb@}@ - such initiation of table elements is correct. Null value will be attributed to 		each element. Compiler may signal lack of initiation values.
	\item Pointers declared in loop `for' work similarly to function inserting new elements in the list. The description of algorithm responsible for this process is in the second point.
	\item Table index is four-bit pattern received after perusal of $K=4$ bits starting with M bit of binary word `data'.
Index value is \verb@(L->data>>M) % @$\Omega$, it's better to replace modulo operation with logical `\verb@AND@' operation
\break
\verb@(L->data>>M) & (@$\Omega$\verb@-1)@, because probably it's five times as fast as remainder of division operator. Such efficiency jump is very important in a loop which decides about the speed of algorithm. If we leave modulo operator, it's possible that compiler will change modulo operator into logical `\verb@AND@' during the process of optimization . Additionally, it's possible to make indexing more optimum:\\
Instead: \verb@in=&tab[ (L->data>>M)&0xf ]@, we can write \\
\verb@in=tab+((L->data>>M)&0xf)@\\
Because `in' is of ``pointer to pointer'' type.
\break
\verb@&tab[index] = &(*(tab+index)) = tab + index@
	\item After inserting node into a new list, pointer `L' will show another element in the list.
	\item In these two lines we decide if we will carry on sorting or if it's high time to merge (sorted) lists.
	If `M' is greater than zero, it means that we haven't looked through all bits yet. This is why we sort all newly created lists in relation to those non-locked through bits -- each separately.
	\item If `M' equals zero it means that we have looked through all bits of sorted numbers. Then, we start the process of list merging (function merge and how it works will be described on the next section).
	\item If the sorted list was empty we return end-maker `P' (NULL) stated in parameter.
If not we  return ``pointer P'' -- indicating the sorted list.
\end{enumerate}

\subsection{Function merge and problem of list merging.}
Function merge was adjusted to the root type. Roots in singly-linked lists can be divided into two groups. Roots of the first group consist of two pointers. One pointer indicates the first element of the list while the other one indicates the last element. The second list contains roots which are the first elements of lists. They are nodes. I have chosen the root which represents the second group approach. Function merge(A,B) merges list `B' with the end of list `A'. Returns address as the first node of a new list. To do that it goes through the whole list `A' and changes bridging of the last link `NULL' into `B'.
Anyone who knows data structures very well (which singly-linked list are) may claim that it would be better to use lists of the first root type. Always, independently of \textbf{\emph{n}} We would make one operations. Using the second type node, however, enables us to make \textbf{\emph{n}} operations. Looking at sorting as a whole the second way will be faster. \emph{Why?} 
In Pbit we split the list more often that we merge it. By using the second type roots we put successive elements to the lists faster.\\
\begin{center}
Definition of function responsible for list merging:\\
\begin{tabular}{|r|l|}
\hline
1 & \verb@Node *merge(Node *A, Node *B){@\\
2 & \textbf{if}\verb@(A==0) @\textbf{return}\verb@ B;@\\
  & \\
3 & \verb@Node *temp=A;@\\
4 & \textbf{while}\verb@(temp->next) temp=temp->next;@\\
5 & \verb@temp->next=B;@\\
6 & \textbf{return}\verb@ A;@\\
  & \verb@}@\\
\hline
\end{tabular}
\end{center}

\begin{enumerate}
	\item Assumption: We insert node `A' at the end list `B'.
  \item If list `A' is empty, we transmit list `B'.
  \item Temporary variable `temp' remembers address of  the first node of list `A'.
  \item We move to the last element of list `A'.
  \item We change bridging of the last element of the list `A' : NULL into `B'.
  \item We return the first element of list `A'.
\end{enumerate}
\emph{Wouldn't it be better to check first it list `B' is empty?\\
Why making loops and going through to the last element of list `A' since in case `B' = NULL we change nothing?} 
No, it wouldn't, because in most cases list `B' will not be empty.
Additional statement ``if'' would make time of list merging worse.
\\
\\
\emph{
You might point out that execution time is $\Theta(n*\frac{M}{K})$, since there are
$\frac{M}{K}$ steps and each of \textbf{n} items is examined in each step.  At worst, the
merging takes another \textbf{n} steps to get to the end of each list.
}
\\
Of course no! For example include counter on merge function:\\
\begin{tabular}{|l|}
\hline
\verb@global@ \textbf{unsigned} \emph{count}\verb@;@\\
\\
\verb@Node *merge(Node *A, Node *B){@\\
\textbf{if}\verb@(A==0) @\textbf{return}\verb@ B; //we do nothing@\\
\emph{count}\verb@++; //first node (maybe only)@\\
\verb@Node *temp=A;@\\
\textbf{while}\verb@(temp->next){@\\
\verb@temp=temp->next;@\\
\emph{count++}\verb@; //n-node of list@\\
\verb@}@\\
\verb@temp->next=B;@\\
\textbf{return}\verb@ A;@\\
\verb@}@\\
\hline
\end{tabular}
\parbox[c]{36ex}{
We group list elements till the last key. When the last key is looked
through, all we have to do is to merge sorted fragments of the list.
Every few, maximum $\frac{M}{K}$, recurrent calls sorted fragment of the list is
(recursion) merged. Pbit by turns merges and splits the lists.\break
\emph{It's ($n*\frac{M}{K}$[split]) \textbf{+} n[merge]}\break
\emph{not  ($n*\frac{M}{K}$[split]) \textbf{*} n[merge]}\break
\textbf{Always \emph{count} equal \emph{n}}\break
Only that it is linear and non-extensive.
}
\newline

\section{Differences between sorting in descending and ascending order}

\textbf{In case of sorting in decreasing order} ($10 \rightarrow 9 \rightarrow 8 \rightarrow \ldots \rightarrow$ NULL)
 lists are merged from the root which represents the smallest elements till the root which represents the greatest ones and the greatest element (node address) is returned. 
\newline
\textbf{In case of sorting in ascending order} ($8 \rightarrow 9 \rightarrow 10 \rightarrow \ldots \rightarrow$ NULL)
 lists are merged from the greatest to the smallest one. The smallest element is returned.
\newline
\begin{center}
\begin{tabular}{|r|l|}
\hline	
 1& \verb@Node *Pbit(Node *L, @\textbf{unsigned}\verb@ M, Node *P=NULL){@\\
 2& \textbf{if}\verb@(L){@\\
 3& \hspace{7 ex}\verb@M-=K;@\\
 4& \hspace{7 ex}\verb@Node *tab[@$\Omega$\verb@]={NULL};@\\
  & \\
 5& \hspace{7 ex}\textbf{for}\verb@(Node *i,**in; i=L; i->next=*in,*in=i){@\\
 6& \hspace{7 ex}\verb@in=tab + ((L->data>>M) & (@$\Omega$\verb@-1));@\\
 7& \hspace{7 ex}\verb@L=L->next;@\\
  & \hspace{7 ex}\verb@}@\\
  & \\
 8& \hspace{7 ex}\textbf{if}\verb@(M) @\textbf{for}\verb@(@\textbf{int}\verb@ i=0; i<@$\Omega$\verb@; i++) P=Pbit(tab[i], M, P);@\\
 9& \hspace{7 ex}\textbf{else for}\verb@(@\textbf{int}\verb@ i=0; i<@$\Omega$\verb@; i++) P=merge(tab[i], P);@\\
  & \hspace{7 ex}\verb@}@\\
  & \\
10& \textbf{return}\verb@ P;@\\
  & \verb@}@\\
\hline	
\end{tabular}	
\end{center}
(8) and (9) line in sorting code undergo change.
\newline
\begin{center}
\textbf{In case of sorting in decreasing order:}
\end{center}
\begin{enumerate}
	\item [8.] \textbf{if}\verb@(M) @\textbf{for}\verb@(@\textbf{int}\verb@ i=0; i<@$\Omega$\verb@; i++) P=Pbit(tab[i], M, P);@
	\item [9.] \textbf{else for}\verb@(@\textbf{int}\verb@ i=0; i<@$\Omega$\verb@; i++) P=merge(tab[i], P);@
\end{enumerate}
We begin to split and merge list from tab[$0$] (list containing the smallest elements) and end with the last list (containing the greatest elements).\break Pointer `P' indicates the node with the highest value.
\begin{center}
\textbf{In case of sorting in ascending order:}
\end{center}
\begin{enumerate}
	\item [8.] \textbf{if}\verb@(M) @\textbf{for}\verb@(@\textbf{int}\verb@ i=(@$\Omega-1$\verb@); i>=0; i--) P=Pbit(tab[i], M, P);@
	\item [9.] \textbf{else for}\verb@(@\textbf{int}\verb@ i=(@$\Omega-1$\verb@); i>=0; i--) P=merge(tab[i], P);@
\end{enumerate}
We begin to split and merge lists from tab[$\Omega-1$] (list containing the greatest elements), and end with the last list (containing the smallest elements).\break Pointer `P' indicates the node with the lowest value.

\section{Negative numbers sorting}
So far we here discussed integer non-negative numbers sorting. To sort the list containing numbers ``with sign'' we use not one but two call a function Pbit. This is why sorting function should require information if the sorted numbers are ``with sign'' or unsigned. It should also take information how to sort - in ascending or decreasing order.
\begin{center}
\verb@Node *Pbit(Node *L, @\textbf{int}\verb@ sign_numbers=1, @\textbf{int}\verb@ sort_decreasing=0);@
\end{center}
We assume that marker \verb@sign_numbers = 1@ and \verb@sort_decreasing = 0@.
\begin{center}
\verb@sign_numbers = 1@\\
It means that we sort numbers ``with sign''.\\
\end{center}
The problem with `\verb@sign_numbers@' definition we may leave Pbit function or use ``Run-Time Type Identification'' from \verb@typeinfo.h@ library.
\begin{center}
\begin{tabular}{|l|}
\hline
\verb@template<class type> inline int sign_numbers(type){@\\
\verb@return (type)((type)1<<((sizeof(type)<<3)-1))<0;@\\
\verb@}@\\
\hline
\end{tabular}
\break Only for integer the function is able to do!\\
\end{center}

\begin{center}
\verb@sort_decreasing = 0@\\
It means that we sort in ascending order. \break
Each value than `$0$' indicates sorting in ascending order.
\end{center}

\begin{center}
\flushleft{Pbit written in pseudo-code}\break
\begin{tabular}{|l|}
\hline
\small{\textbf{if}(}\verb@sign_numbers@\small{)}\{\\
\small{Sorts numbers `with sign' (positive and negative)}\\
\small{\textbf{if}(}\verb@sort_decreasing@\small{) Sorts in decreasing order}\\
\small{\textbf{else} Sorts in ascending order}\\
\small{\}}\\
\small{\textbf{else}\{}\\
\small{Sorts numbers unsigned (non-negative)}\\
\small{\textbf{if}(}\verb@sort_decreasing@\small{) Sorts in decreasing order}\\
\small{\textbf{else} Sorts in ascending order}\\
\small{\}}\\
\hline
\end{tabular}
\parbox[c]{30ex}{
Numbers ``with sign'' and non-negative (unsigned) are recorded in a different way.
For representing numbers ``with sign'' so called ``binary complement'' is use.
To understand fragment of the code responsible for sorting numbers ``with sign'' we must only know that the most 
}
\break
\parbox[c]{\textwidth}{
important bits of negative numbers are always greater than the most important bits of positive numbers.
This is why while sorting negative numbers using Pbit, which sorts in decreasing order, in fact receive ascending list. There is a complete Pbit code written in C++ on another section.
}
\end{center}

\section{Complete Pbit implementation in C++}
\begin{center}
\begin{tabular}{|r|l|}
\hline
 1& \verb@Node *Pbit(Node *L,@\textbf{int}\verb@ sign_numbers,@\textbf{int}\verb@ sort_decreasing);@\\
 2& \textbf{if}\verb@(L==0) @\textbf{return}\verb@ 0;@\\
 3& \textbf{unsigned}\verb@ size=sizeof(L->data)<<3;@\\
  & \verb@//sign_numbers = (1)-sort numbers `with sign' (negative)@\\
  & \verb@//               (0)-unsigned (non-negative)@\\
 4& \textbf{if}\verb@(sign_numbers){@\\
 5& \verb@Node *L_unsigned = 0, *L_negative = 0;@\\
  & \verb@//split list into negative and@\\ 
  & \verb@//non-negative (unsigned) numbers@\\
 6& \textbf{for}\verb@(@\textbf{register}\verb@ Node *temp; temp = L; ){@\\
 7& \verb@L = L->next;@\\
 8& \textbf{if}\verb@(temp->data < 0){@\\
 9& \verb@temp->next = L_negative; L_negative = temp;@\\
10& \verb@}@\textbf{else}\verb@{ temp->next = L_unsigned; L_unsigned = temp; }@\\
  & \verb@}@\\
11& \textbf{return}\verb@ (sort_decreasing)?@\\
12& \verb@Pbit_decreasing(L_unsigned, size, @\\
  & \hspace{20 ex}\verb@Pbit_decreasing(L_negative, size) ):@\\
13& \verb@Pbit_ascending(L_negative, size, @\\
  & \hspace{18 ex}\verb@Pbit_ascending(L_unsigned, size) );@\\
  & \verb@}@\\
14& \textbf{else return}\verb@ (sort_decreasing)? Pbit_decreasing(L, size):@\\
15& \hspace{37 ex}\verb@Pbit_ascending(L, size);@\\
  & \verb@}@\\
\hline
\end{tabular}
\verb@Pbit_decreasing@ -- we sort list in decreasing order\break (returns pointer to the greatest element).\\
\verb@Pbit_ascending@ -- we sort list in ascending order\break (returns pointer to the smallest element).
\end{center}
\begin{enumerate}
	\item [2.] We check if sorted list `L' is not empty.
	\item [3.] We determine the size of sorted element in bits.
	\item [4.] If (logical) variable \verb@sign_numbers@ if different than zero, it means that sorted elements are signed.
	\item [5.] We create two pointers which will become the roots of new lists; list representing negative numbers and list representing non-negative (unsigned) numbers.
	\item [6-10.] We split lists, list roots represent nodes (5).
	\item [11.] If variable \verb@sort_decreasing@ is\ldots 
	\item [12.] different than zero we sort in decreasing order ($10 \rightarrow 9 \rightarrow 8 \rightarrow \ldots $),
	\item [13.] if not we sort in ascending order ($8 \rightarrow 9 \rightarrow 10 \rightarrow \ldots $).
	\item [14.] Variable \verb@sign_numbers@ is equal to zero. It means that we sort numbers ``unsigned'' (non-negative).
	\item [14-15.] If variable \verb@sort_decreasing@ in different than zero we sort in decreasing order, if not we sort in ascending order.
\end{enumerate}

\section{Comparison of list sorting algorithms}
\textbf{QuickerSort}~\cite{Scowen} is characterised by very small O(nlgn) coefficient. The number of other algorithm operations is not significantly greater than number of comparison. QuickerSort is based on list division in relation to the first node (root). We divide the list into elements which are equal, greater or less than the chosen node. Next, the ``biggest'' (by mean of value) and the ``smallest'' lists are recurrently sorted. Its most serious disadvantages are instability and cost O($n^2$) in worst case behavior. In storage complexity we must take into account memory for stack and heap implementation. As far as worst case behaviour is concerned, the depth recursion is $n-1$. 
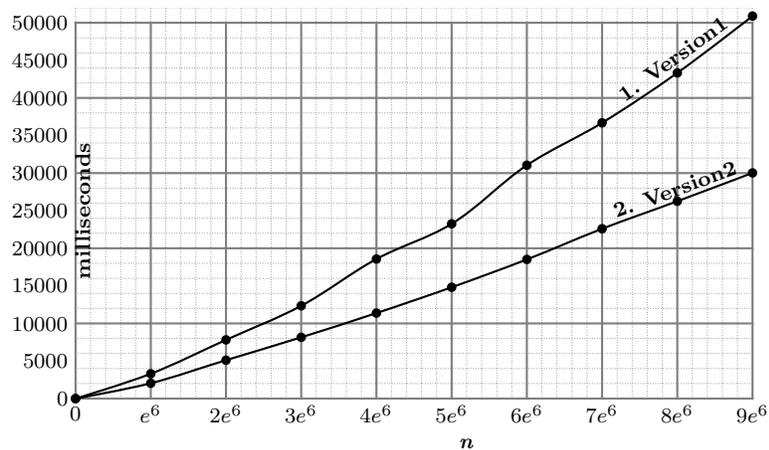
\begin{small}
\begin{center}
\begin{tabular}{@{} c | c @{}}
Version 1& Version 2\\
\hline
\begin{tabular}{| p{160pt} @{} }
\verb@Node *merge(Node *t, Node *P){@\\
\verb@if(t==0) return P;@\\
\\
\verb@Node *tmp = t;@\\
\verb@while(tmp->next) tmp=tmp->next;@\\
\verb@tmp->next=P;@\\
\\
\verb@return t;@\\
\verb@}@\\
\\
\\
\verb@Node *qs(Node *L, Node *P=0){@\\
\verb@if(L==0) return P;@\\
\\
\verb@Node *L1=0, *L2=0, *L3=0;@\\
\verb@L2=L; L=L->n; L2->next=0;@\\
\\
\verb@for(Node *i;i=L;){@\\
\verb@L=L->next;@\\
\verb@if(L2->data > i->data)@\\
\verb@{ i->next=L1;  L1=i; }@\\
\verb@else if(L2->data == i->data)@\\
\verb@{ i->next=L2;  L2=i; }@\\
\verb@else@\\
\verb@{ i->next=L3;  L3=i; }@\\
\verb@}@\\
\\
\verb@return@\\ 
\verb@merge(qs(L3),merge(L2,qs(L1)));@\\
\verb@}@\\
\end{tabular} 

&

\begin{tabular}{ @{} p{160pt} |} 
\verb@Node *qs(Node *L, Node **P=0){@\\
\verb@if(L==0) return 0;@\\
\verb@if(L->next == 0) return (*P = L);@\\
\\
\verb@Node *L1=0, *L2=L, *L3=0;@\\
\verb@Node *P1=0, *P2=L, *P3=0;@\\
\\
\verb@L=L->next;@\\
\verb@L2->next=0;@\\
\verb@for(Node *i;i=L;){@\\
\verb@L=L->next;@\\
\verb@if(L2->data < i->data)@\\
\verb@{ i->next=L1; L1=i; }@\\
\verb@else if(L2->data == i->data)@\\
\verb@{ i->next=L2; L2=i; }@\\
\verb@else@\\
\verb@{ i->next=L3; L3=i; }@\\
\verb@}@\\
\\
\verb@L3=qs(L3, &P3);@\\
\verb@if(L3) P3->next=L2;@\\
\verb@else L3=L2;@\\
\verb@if(L1){@\\ 
\verb@P2->next=qs(L1, &P1);@\\
\verb@if(P) *P=P1;@\\
\verb@}@\\
\verb@else if(P) *P=P2;@\\
\verb@return L3;@\\
\verb@}@\\
\\
\end{tabular}
\\
\hline
\end{tabular}
\end{center}
\end{small}

\begin{center}
\begin{footnotesize}
\begin{pspicture*}(-1,-1)(10,5.2)
\psgrid[gridcolor=gray, subgriddots=6](-1,-0.6)(0,0)(9,5.1)

\rput[b]{36}(8, 4.45){\textbf{1. Version1}}
\listplot[plotstyle=curve, showpoints=true]{
0 0
1 0.3305
2 0.7811
3 1.2348
4 1.8567
5 2.3253
6 3.1045
7 3.6693
8 4.3333
9 5.0893
}

\rput[b]{20}(8, 2.7){\textbf{2. Version2}}
\listplot[plotstyle=curve, showpoints=true]{
0 0
1 0.2043
2 0.5107
3 0.8152
4 1.1376
5 1.4811
6 1.8516
7 2.2593
8 2.6257
9 3.0025
}

\rput[l](5,-0.6){ \textbf{\emph{n}}}
\rput[c](0,-0.2){$0$}
\rput[c](1,-0.2){$e^6$}
\rput[c](2,-0.2){$2e^6$}
\rput[c](3,-0.2){$3e^6$}
\rput[c](4,-0.2){$4e^6$}
\rput[c](5,-0.2){$5e^6$}
\rput[c](6,-0.2){$6e^6$}
\rput[c](7,-0.2){$7e^6$}
\rput[c](8,-0.2){$8e^6$}
\rput[c](9,-0.2){$9e^6$}

\rput[r](-0.1, 0){$0$}
\rput[r](-0.1, 0.5){$5000$}
\rput[r](-0.1, 1){$10000$}
\rput[r](-0.1, 1.5){$15000$}
\rput[r](-0.1, 2){$20000$}
\rput[r](-0.1, 2.5){$25000$}
\rput[r](-0.1, 3){$30000$}
\rput[r](-0.1, 3.5){$35000$}
\rput[r](-0.1, 4){$40000$}
\rput[r](-0.1, 4.5){$45000$}
\rput[r](-0.1, 5){$50000$}
\rput[t]{90}(0, 2.5){\textbf{milliseconds}}

\end{pspicture*}

\begin{tabular}{|c|c|c|c|c|c|c|c|c|c|}
\hline
\textbf{\emph{n}} & $e^{6}$ &	$2e^{6}$	&$3e^{6}$&	$4e^{6}$&	$5e^{6}$&	$6e^{6}$&	$7e^{6}$&	$8e^{6}$&	$9e^{6}$\\
\hline
1. & 3305 &	7811 &	12348	& 18567	& 23253	& 31045	& 36693	& 43333	& 50893\\
\hline
2. & 2043 &	5107 &	8152&	11376&	14811&	18516&	22593&	26257	&30025\\
\hline
\end{tabular}
\end{footnotesize}
\end{center}
\begin{enumerate}
	\item QuickerSort (with outside the merge function) Version1.
	\item QuickerSort (without the merge function) Version2\footnote{To make a comparison I used more efficient version2.}.
\end{enumerate}
\textbf{MergeSort}~\cite{MergeSort} like QuickerSort is based on element comparison. The best description of MergeSort is based on recursion. Algorithm divides list into two lists of the some size. Next, both lists are sorted (MergeSort) separately. To finish the process of sorting original n element list, two sorted halves are merged. Unfortunately, recurrent implementation of algorithm does not sort ``in situ'', so we need twice as big memory as occupied by unsorted data. On my hardware version1 of algorithm~\cite{Alf} stopped list sorting which consisted of over million elements -- it overflow the stack. It considerably restricts usefulness of ``advanced recurrent'' implementation of algorithm. This is why I used to comparision \emph{\textbf{partly iterative}}(version2), more efficient MergeSort implementation.
\begin{small}
\begin{center}
\begin{tabular}{@{} c | c @{}}
Version 1& Version 2\\
\hline
\begin{tabular}{| p{160pt} @{} } 
\verb@Node *MergeSort(Node *);@\\
\verb@Node *merge(Node *, Node *);@\\
\verb@Node *split(Node *);@\\
\\
\\
\verb@Node *MergeSort(Node *list){@\\
\verb@if(list==0) return 0;@\\
\verb@if(list->next==0) return list;@\\
\verb@Node *SecondList=split(list);@\\
\verb@return merge(MergeSort(list),@\\
\verb@MergeSort(SecondList));@\\
\verb@}@\\
\\
\\
\verb@Node *merge(Node *list1,@\\
\verb@Node *list2){@\\
\verb@if(list1==0) return list2;@\\
\verb@if(list2==0) return list1;@\\

\verb@ if(list1->data <= list2->data){@\\
\verb@ list1->next=@\\
\verb@  merge(list1->next, list2);@\\
\verb@ return list1;@\\
\verb@ }@\\
\verb@ else{@\\
\verb@ list2->next=@\\
\verb@  merge(list1,list2->next);@\\
\verb@ return list2;@\\
\verb@ }@\\
\\
\verb@}@\\
\\
\\
\verb@Node *split(Node *list){@\\
\verb@if(list==0) return 0;@\\
\verb@if(list->next==0) return 0;@\\
\\
\verb@Node *pSecondCell=list->next;@\\
\verb@list->next=pSecondCell->next;@\\
\verb@pSecondCell->next=@\\
\verb@ split(pSecondCell->next);@\\
\verb@return pSecondCell;@\\
\verb@}@\\
\\ \\ \\ \\ \\ \\
\end{tabular}

&

\begin{tabular}{ @{} p{160pt} |}
\verb@Node *MergeSort(Node *);@\\
\verb@Node *merge(Node *, Node *);@\\
\verb@Node *split(Node *);@\\
\\
\\
\verb@Node *MergeSort(Node *list){@\\
\verb@if(list==0) return 0;@\\
\verb@if(list->next==0) return list;@\\
\verb@Node *SecondList=split(list);@\\
\verb@return merge(MergeSort(list),@\\ 
\verb@MergeSort(SecondList));@\\
\verb@}@\\
\\
\\
\verb@Node *merge(Node *list1,@\\
\verb@Node *list2){@\\
\verb@	if(list1 == 0) return list2;@\\
\verb@	if(list2 == 0) return list1;@\\
\\
\verb@	Node *list = 0, *last = 0;@\\
\verb@	while(list1 && list2){@\\
\verb@		Node* p;@\\
\verb@		if(list1->data <= list2->data){@\\
\verb@			p = list1; list1 = list1->next;@\\
\verb@		}else{@\\
\verb@			p = list2; list2 = list2->next;@\\
\verb@		}@\\
\verb@		if(list) last = last->next = p;@\\
\verb@		else last = list = p;@\\
\verb@	}@\\
\verb@	if(list1) last->next = list1;@\\
\verb@	else last->next = list2;@\\
\verb@	return list;@\\
\verb@}@\\
\\
\\
\verb@Node *split(Node *list)@\\
\verb@{@\\
\verb@	Node* list2 = 0;@\\
\verb@	while(list && list->next){@\\
\verb@		Node* w = list->next->next;@\\
\verb@		list->next->next = list2;@\\
\verb@		list2 = list->next;@\\
\verb@		list->next = w;@\\
\verb@		list = list->next;@\\
\verb@	}@\\
\verb@	return list2;@\\
\verb@}@\\
\end{tabular} 
\\
\hline
\end{tabular}
\end{center}
Unfortunately, recursion MergeSort implementation overflows the stack by sorting over-million list.
For such list it obtained time 6569 (version2: 4016) milliseconds.\\
\end{small}
\\
I proposed \textbf{Psort} instead of QuickerSort to lists sorting. Algorithm divides the list into three sub-lists in relation to two nodes, into elements less than the first node, greater than the first node but less than the second node and greater than the second node. It's excellent for sorting short lists.
\begin{small}
\begin{center}
\begin{tabular}{| l |}
\hline
\verb@Node *Psort(register Node *L, Node *P=0){@\\
\verb@if(L==0) return P;@\\
\verb@if(L->next==0) return (L->next=P,L);@\\
\\
\verb@Node *P2, *L2=0;@\\
\verb@{@\\
\verb@Node *i,@\\
\verb@*tmp=(L->data > L->next->data? P2=L,L=L->next : P2=L->next)->next;@\\
\\
\verb@for(P2->next=L->next=0; i=tmp; ){@\\
\verb@tmp=tmp->next;@\\
\verb@if(i->data < L->data){ i->next=L->next; L->next=i; }@\\
\verb@else if(i->data > P2->data){ i->next=L2; L2=i; }@\\
\verb@else{ i->next=P2->next; P2->next=i; }@\\
\verb@}@\\
\\
\verb@}@\\
\\
\verb@L->next=Psort(L->next,P);@\\
\verb@P2->next=Psort(P2->next,L);@\\
\verb@return Psort(L2,P2);@\\
\verb@}@\\
\hline
\end{tabular}
\end{center}
\end{small}
\textbf{Psort2} is modification of \textbf{Psort}. I implemented it for sorting long lists with often repeated value. Wegner~\cite{Weg} describes similar ternary partitioning algorithm more efficient than QuickSort.
\begin{small}
\begin{center}
\begin{tabular}{| l |}
\hline
\verb@Node *Psort2(Node *L, Node *P=0){@\\
\verb@if(L==0) return P;@\\
\verb@if(L->next==0) return (L->next=P,L);@\\
\\
\verb@Node *P2, *L2=0;@\\
\verb@{//---2@\\
\verb@Node *PP;@\\
\\
\verb@{//---1@\\
\verb@Node *LL;@\\
\\
\verb@{//---0@\\
\verb@ Node *i, @\\
\verb@ *tmp=(L->data > L->next->data? P2=L,L=L->next : P2=L->next)->next;@\\
\\
\verb@ for(PP=P2, LL=L, P2->next=L->next=0; i=tmp; ){//_for@\\
\verb@  tmp=tmp->next;@\\
\verb@  if(i->data < L->data){ i->next=L->next; L->next=i; }@\\
\verb@  else if(i->data > P2->data){ i->next=L2; L2=i; }@\\
\verb@  else{//_else1@\\
\verb@   if(i->data == L->data){i->next=LL; LL=i;}@\\
\verb@   else if(i->data == P2->data){i->next=PP; PP=i;}@\\
\verb@   else{i->next=P2->next; P2->next=i;}@\\
\verb@  }//else1_@\\
\verb@ }//for_@\\
\verb@}//---0@\\
\\
\verb@L->next=Psort2(L->next,P);@\\
\verb@L=LL;@\\
\verb@}//---1@\\
\\
\verb@P2->next=Psort2(P2->next,L);@\\
\verb@P2=PP;@\\
\verb@}//---2@\\
\\
\verb@return Psort2(L2,P2);@\\
\verb@}@\\
\hline
\end{tabular}
\end{center}
\end{small}
\textbf{Pbit} sorts by grouping nodes. It belongs to the group of algorithms which I worked out and which uses ``end marker'' `P'. It is linear sorting algorithm with coefficient dependent on the length of pattern `\textbf{\emph{K}}'. Conventionally pattern was set up for four bits. It makes sorting both short and long lists effective. It also has influence on very little need for memory, depending on size of sorted elements and not their number (it's non-extensive). Each programmer may change the length of the pattern and adept Pbit for his use. Such problems as ``worst case behaviour'' do not occur in case of Pbit. Data base programmers don't have to create a few lists, attribute them to particular data in structure and sort every list separately owing to stability of the algorithm. Above fact show why Pbit is considered not only as the fastest but also the most suitable for lists sorting.
\rule{\textwidth}{0.1pt}

\begin{footnotesize}
Tests were carried out on computer equipped with processor AMD Athlon XP 1800+ (1533 Mhz).
For needs of tests I filled the list with pseudo-random numbers obtained from function rand(), 
\verb@RAND_MAX=0x7FFF@, from standard library C.
Each time I initiated current time into pseudo-random numbers generator.
Every test was carried out ten times, arithmetic mean of obtained results was written in the table.
\end{footnotesize}

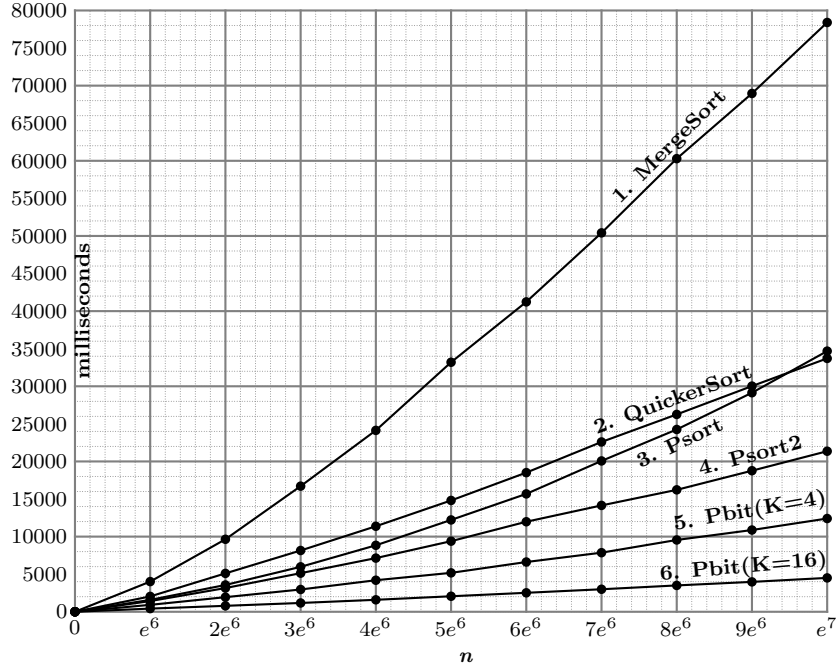
\begin{center}
\begin{footnotesize}
\begin{pspicture*}(-1,-1)(11,8.2)
\psgrid[gridcolor=gray, subgriddots=6](-1,-0.6)(0,0)(10,8)

\rput[b]{45}(8, 6.1){\textbf{1. MergeSort}}
\psline[linearc=0.25, showpoints=true](0,0)(1, 0.4016)(2, 0.9644)(3, 1.6714)(4, 2.4134)(5, 3.3198)(6, 4.1229)(7, 5.0412)(8, 6.0287)(9, 6.8959)(10, 7.8403)

\rput[b]{20}(8, 2.7){\textbf{2. QuickerSort}}
\psline[linearc=0.25, showpoints=true](0,0)(1, 0.2043)(2, 0.5107)(3, 0.8152)(4, 1.1376)(5, 1.4811)(6, 1.8516)(7, 2.2593)(8, 2.6257)(9, 3.0025)(10, 3.3710)

\rput[t]{25}(8, 2.35){\textbf{3. Psort}}
\psline[linearc=0.25, showpoints=true](0,0)(1, 0.1582)(2, 0.3565)(3, 0.5979)(4, 0.8842)(5, 1.2218)(6, 1.5683)(7, 2.0069)(8, 2.4254)(9, 2.9142)(10, 3.4680)
	
\rput[b]{15}(9, 2){\textbf{4. Psort2}}
\psline[linearc=0.25, showpoints=true](0,0)(1, 0.1422)(2, 0.3155)(3, 0.5137)(4, 0.7141)(5, 0.9393)(6, 1.1977)(7, 1.4150)(8, 1.6223)(9, 1.8767)(10, 2.1361)

\rput[b]{10}(9, 1.2){\textbf{5. Pbit(K=4)}}
\psline[linearc=0.25, showpoints=true](0,0)(1, 0.0921)(2, 0.1943)(3, 0.2964)(4, 0.4196)(5, 0.5178)(6, 0.6619)(7, 0.7862)(8, 0.9553)(9, 1.0876)(10, 1.2407)

\rput[b]{5}(8.9, 0.45){\textbf{6. Pbit(K=16)}}
\psline[linearc=0.25, showpoints=true](0,0)(1, 0.0421)(2, 0.0791)(3, 0.1171)(4, 0.1592)(5, 0.2063)(6, 0.2524)(7, 0.2995)(8, 0.3505)(9, 0.3976)(10, 0.4506)

\rput[l](5,-0.6){ \textbf{\emph{n}}}
\rput[c](0,-0.2){$0$}
\rput[c](1,-0.2){$e^6$}
\rput[c](2,-0.2){$2e^6$}
\rput[c](3,-0.2){$3e^6$}
\rput[c](4,-0.2){$4e^6$}
\rput[c](5,-0.2){$5e^6$}
\rput[c](6,-0.2){$6e^6$}
\rput[c](7,-0.2){$7e^6$}
\rput[c](8,-0.2){$8e^6$}
\rput[c](9,-0.2){$9e^6$}
\rput[c](10,-0.2){$e^7$}

\rput[r](-0.1, 0){$0$}
\rput[r](-0.1, 0.5){$5000$}
\rput[r](-0.1, 1){$10000$}
\rput[r](-0.1, 1.5){$15000$}
\rput[r](-0.1, 2){$20000$}
\rput[r](-0.1, 2.5){$25000$}
\rput[r](-0.1, 3){$30000$}
\rput[r](-0.1, 3.5){$35000$}
\rput[r](-0.1, 4){$40000$}
\rput[r](-0.1, 4.5){$45000$}
\rput[r](-0.1, 5){$50000$}
\rput[r](-0.1, 5.5){$55000$}
\rput[r](-0.1, 6){$60000$}
\rput[r](-0.1, 6.5){$65000$}
\rput[r](-0.1, 7){$70000$}
\rput[r](-0.1, 7.5){$75000$}
\rput[r](-0.1, 8){$80000$}

\rput[t]{90}(0, 4){\textbf{milliseconds}}
\end{pspicture*}

\begin{tabular}{|c|c|c|c|c|c|c|c|c|c|c|}
\hline
\textbf{\emph{n}} & $e^{6}$ &	$2e^{6}$	&$3e^{6}$&	$4e^{6}$&	$5e^{6}$&	$6e^{6}$&	$7e^{6}$&	$8e^{6}$&	$9e^{6}$	&$e^{7}$\\
\hline
1. & 4016 &	9644 &	16714	& 24134	& 33198	& 41229	& 50412	& 60287	& 68959 & 78403\\
\hline
2. & 2043 &	5107 &	8152&	11376&	14811&	18516&	22593&	26257	&30025&	34410\\
\hline
3. & 1582 & 3565	& 5979	&8842	&12218&	15683	&20069&	24254	&29142&	34680\\
\hline
4. & 1422 & 3155	& 5137	&7141&	9393&	11977&	14150&	16223	&18767&	21361\\
\hline
5. & 921	& 1943 & 2964	&4196	&5178	&6619	&7862	&9553	&10876	&12407\\
\hline
6. & 421	& 791	& 1171	&1592&	2063&	2524&	2995	&3505&	3976	&4506\\
\hline
\end{tabular}
\begin{flushleft}
Unfortunately, recursion MergeSort implementation overflows the stack by sorting over-million list.
For such list it obtained time 6569 milliseconds. I used iterative MergeSort implementation in the test.
\end{flushleft}
\end{footnotesize}
For ten-million (long) list Pbit sorting with sixteen-bit pattern is more than
\textbf{seventeen times faster} than the most popular MergeSort!

\end{center}

\subsection{Point on axis n (number of elements) in relation to linear-logarithmic order}
Probably each linear algorithm has some point on axis n (number of elements) in relation to which it is slower than linear-logarithmic algorithm. Let's check from which point on axis O(n) Pbit is slower than QuickerSort O(nlgn).\\
\\
For bit pattern which equals four we sort n four-byte natural numbers:\\
For $K=4$ bits, $M=32$ bits\\
\\
$\Theta(8n)$ will be faster than $\Theta(nlog_{2}n)$ for $n>256$ elements.\\
However we cannot compare only coefficients since the most important operations of algorithms differ in time. In respect of time QuickerSort differs from Pbit on statement responsible list division:

\begin{center}
\begin{tabular}{c|c}
QuickerSort & Pbit\\
\hline
\parbox[t]{35ex}{if(data$>$x)\\else if(data$==$x)\\else} &
\parbox[t]{35ex}{tab[(data$>>$M) \& 0xF]\\0xF = ($\Omega-1$) -- example constant}\\
\hline
\parbox[t]{35ex}{At the average n elements will be compared $1.5nlog_{2}n$ times} & 
\parbox[t]{35ex}{In case of Pbit n elements\break (relations) will be shifted\break $(M/K)*n$ times}\\
\hline
\end{tabular}

\parbox{70ex}{
Now we will analyse time consumption of operations. Because all ticks of the clock are the same, we omit them and give only mean time in nanosecond.
}

\begin{tabular}{c|c}
\hline
\parbox[t]{35ex}{Operation ``if'' in C requires $20$ nanosecond on the average} & 
\parbox[t]{35ex}{Operations of bit shift, logical `AND' and reference to table element take in total 45 nanoseconds on the average in C}\\
\hline
\parbox[t]{35ex}{n elements sorted in \mbox{$20*1.5*nlog_{2}n = 30nlog_{2}n$} nanoseconds} & 
\parbox[t]{35ex}{n elements sorted in $45*8n = 360n$ nanoseconds}\\
\hline
\end{tabular}
\parbox{25ex}{
$30nlog_{2}n > 360n, n>0$\\
$nlog_{2}n > 12n$\\
$log_{2}n > 12$\\
$n > 4096$\\
}

\parbox{70ex}{For $n > 4096$ elements Pbit which splits thirty two-bit numbers with four-bit pattern will be faster.}
\end{center}
It does not mean, however, that we will see the differences while sorting less than $4096$ elements. Modern computers will need only one hundredth of second to sort (linear-logarithmic) even ten thousand elements. \textbf{\emph{It means that we should always use Pbit for lists sorting.}}
For long lists we can enlarge the length of the pattern. It will make the coefficient smaller and consequently the algorithm will be faster.
For example, for $K=8$, coefficient equals four, which in result will cause efficiency jump and lessen the number of nodes, which linear Pbit will always be faster than linear-logarithmic algorithms to sixty four elements!

\section{The length of bit pattern}
For longer pattern algorithm will require more memory.\\
Maximum amount of occupied storage can be expressed by a formula:\break
$T = (\Omega*4 + 3*4) * \frac{M}{K} + 2*4$ [byte]\\
If $K=4$, algorithm will need $616$ bytes to sort four-byte numbers.\break
Whereas for $K=8$, algorithm will require as much as $4152$ bytes.

\begin{center}
\begin{tabular}{| l |}
\hline
\textbf{\emph{Constants:}} $K=8$, $\Omega=256$, hex($\Omega-1$) = hex($256-1$) = hex($255$) = 0xFF
\\
\\
\verb@Node *Pbit(Node *L, unsigned M, Node *P=NULL){@\\
\verb@ if(L){@\\ 
\verb@          M-=@\textbf{\emph{8}}\verb@;@\\
\verb@          Node *tab[@\textbf{\emph{256}}\verb@]={NULL};@\\
\\
\verb@          for(Node *i,**in; i=L; i->n=*in,*in=i){@\\
\verb@          in=tab + ( (L->data>>M) & @\textbf{\emph{0xFF}}\verb@ );@\\
\verb@          L=L->next;@\\
\verb@          }@\\
\\
\verb@          if(M) for(int i=0; i<@\textbf{\emph{256}}\verb@; i++)  P=Pbit(tab[i], M, P);@\\
\verb@          else   for(int i=0; i<@\textbf{\emph{256}}\verb@; i++)  P=merge(tab[i], P);@\\
\verb@         }@\\
\\
\verb@ return P;@\\
\verb@}@\\

\\
\hline
\end{tabular}
\end{center}

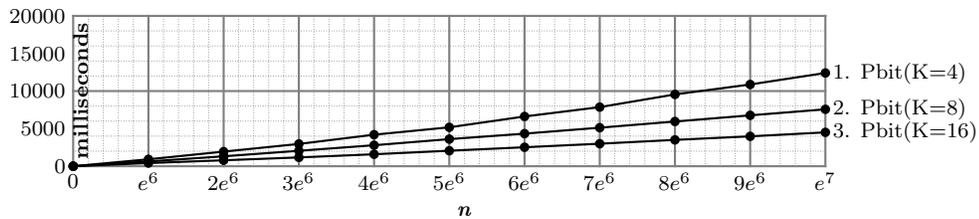
\begin{flushleft}
\begin{footnotesize}
\begin{pspicture*}(-1,-1)(12,2.2)
\psgrid[gridcolor=gray, subgriddots=6](-1,-0.6)(0,0)(10,2)

\rput[l](10.1 ,1.2407){1. Pbit(K=4)}
\psline[linearc=0.25, showpoints=true](0,0)(1, 0.0921)(2, 0.1943)(3, 0.2964)(4, 0.4196)(5, 0.5178)(6, 0.6619)(7, 0.7862)(8, 0.9553)(9, 1.0876)(10, 1.2407)

\rput[l](10.1 ,0.7571){2. Pbit(K=8)}
\psline[linearc=0.25, showpoints=true](0,0)(1, 0.0631)(2, 0.1322)(3, 0.2053)(4, 0.2794)(5, 0.3605)(6, 0.4336)(7, 0.5127)(8, 0.5948)(9, 0.6770)(10, 0.7571)

\rput[l](10.1 ,0.4506){3. Pbit(K=16)}
\psline[linearc=0.25, showpoints=true](0,0)(1, 0.0421)(2, 0.0791)(3, 0.1171)(4, 0.1592)(5, 0.2063)(6, 0.2524)(7, 0.2995)(8, 0.3505)(9, 0.3976)(10, 0.4506)

\rput[l](5,-0.6){ \textbf{\emph{n}}}
\rput[c](0,-0.2){$0$}
\rput[c](1,-0.2){$e^6$}
\rput[c](2,-0.2){$2e^6$}
\rput[c](3,-0.2){$3e^6$}
\rput[c](4,-0.2){$4e^6$}
\rput[c](5,-0.2){$5e^6$}
\rput[c](6,-0.2){$6e^6$}
\rput[c](7,-0.2){$7e^6$}
\rput[c](8,-0.2){$8e^6$}
\rput[c](9,-0.2){$9e^6$}
\rput[c](10,-0.2){$e^7$}

\rput[r](-0.1, 0){$0$}
\rput[r](-0.1, 0.5){$5000$}
\rput[r](-0.1, 1){$10000$}
\rput[r](-0.1, 1.5){$15000$}
\rput[r](-0.1, 2){$20000$}

\rput[t]{90}(0, 1){\textbf{milliseconds}}

\end{pspicture*}
\end{footnotesize}
\end{flushleft}

\begin{center}
\begin{footnotesize}
\begin{tabular}{|c|c|c|c|c|c|c|c|c|c|c|}
\hline
\textbf{\emph{n}} & $e^{6}$ &	$2e^{6}$	&$3e^{6}$&	$4e^{6}$&	$5e^{6}$&	$6e^{6}$&	$7e^{6}$&	$8e^{6}$&	$9e^{6}$	&$e^{7}$\\
\hline
1.&921	&1943	&2964	&4196	&5178	&6619	&7862	&9553	&10876	&12407\\
\hline
2.&631	&1322	&2053	&2794	&3605	&4336	&5127	&5948	&6770	&7571\\
\hline
3.&421	&791	&1171	&1592	&2063	&2524	&2995	&3505	&3976	&4506\\
\hline
\end{tabular}
\end{footnotesize}
\end{center}

\section{Differences between Pbit and other algorithms sorting with ``\emph{bit~key}''}
\emph{I will compare Pbit with algorithm described by Donald E. Knuth~\cite{Knuth}.}
\\
\\
\textbf{Address counting sort}\\
In case of Pbit there is no worst case. If all elements are identical, it takes far less time to sort the list than for non-uniform distribution of the keys.
\\
\textbf{Position interchange}\\
Position sorting by swamping can be used to sort mainly tables.
Pbit does not look through the both sides of the list.
\\
\textbf{Scatter sort}\\
Recurrent scatter sort, unlike Pbit, has very high proportionality factor.
\\
\textbf{List position sorting}\\
Pbit scatters in relation to the most significant bits, not to the least significant ones.
We scatter list elements till the last bit and lists scattered that way are merged. 
Only sorted of the list are merged.
\\
\textbf{Position Sort in relation to the most significant number}\\
Algorithm spends too much time on small ``chunks''.
Algorithm sorting in relation to the least significant numbers is relatively efficient.
Probably the best compromise was suggested by M.D. MacLaren in 1966~\cite{MacLaren}. He recommended using sorting with the least significant number as first but use it only for most significant numbers. After that sequence will not be completely sorted but almost ordered, so not much inversion will be left there. This is why it is possible to finish sorting by using simple insertion.
Also W. Dobosiewicz~\cite{Dobosiewicz} dealt with algorithm sorting with the most significant number as first until he obtained short subsequences. Unfortunately, in case of non-uniform key arrangement, algorithm has worst case O(nlgn). The work of Dobosiewicz inspired some other scholars to invent new algorithms based on address counting the best known of which is perhaps the scheme worked out by M. Tamminen in 1985~\cite{Tamminen}: 
``Let's assume that all keys are fractions from range $[0..1)$. First, we scatter all N records to $N/8$ baskets, putting key $K$ into basket $KN/8$ (\ldots)''.
\\
\\
\textbf{While looking through described algorithms we can see considerable differences concerning mainly merging of sequence element. 
Pbit merges the lists only when the lists have been sorted.
`` We group list elements till the last key. When the last key is looked through, all we have to do is to merge sorted fragments of the list. Every few, maximum M/K, recurrent calls sorted fragment of the list is (recursion) merged.''
Recursion and the part of code responsible for merging sorted fragments of the list make Pbit very efficient and \emph{innovated} algorithm.}
\\

\begin{center}
\begin{tabular}{|l|}
\hline
\textbf{if}\verb@(M) @\textbf{for}\verb@(@\textbf{int}\verb@ i=0; i<@$\Omega$\verb@; i++) P=Pbit(tab[i], M, P);@\\
\textbf{else for}\verb@(@\textbf{int}\verb@ i=0; i<@$\Omega$\verb@; i++) P=merge(tab[i], P);@\\
\hline
\end{tabular}
\end{center}
\rule{\textwidth}{0.1pt}
\break
\textbf{All algorithms sorting with ``bit key'' are very \emph{alike}.
Very similar are also all algorithms which sort by comparing elements.
However, \emph{comparison implementation} makes QuickSort faster than InsertionSort. 
Just like sorting with ``bit key'' implementation makes\break Pbit the most efficient algorithm which sorts lists.
}
\break
\rule{\textwidth}{0.1pt}

\section{Tables versus list}
What are the most important differences between a table and list?
Firstly, tables have fixed size whereas list size depends on the size of data stored there, enlarged by the size of storage area necessary to keeps pointers for its every element. 
Secondly, to change the order of list elements, it is enough to change whole a few pointers, which is far less expensive than moving whole portion of memory in case of change of table elements sequence. 
Thirdly, it's possible to add new elements to the list and remove elements without changing the position of others.
There differences help draw the conclusion that if changes will take place into data set very often, especially when the number of set members is not predictable, then it is good to create a list in order to store data set. 
\\
Nobody has compared algorithms sorting tables and lists. Why?
Because table elements can be freely looked through.
Nodes in singly-linked lists have pointer only for the next node, which makes looking through the lists difficult. Moreover, it's faster to refer to elements of the table than nodes of the lists.
So far the fastest list sorting algorithm was QuickerSort, much slower than linear-logarithmic tables sorting algorithms. It is good to choose tables when we want to have quick access to data and when data were to be sorted.
For the first time in the history of sorting there this can compete with the fastest algorithms tables sorting!
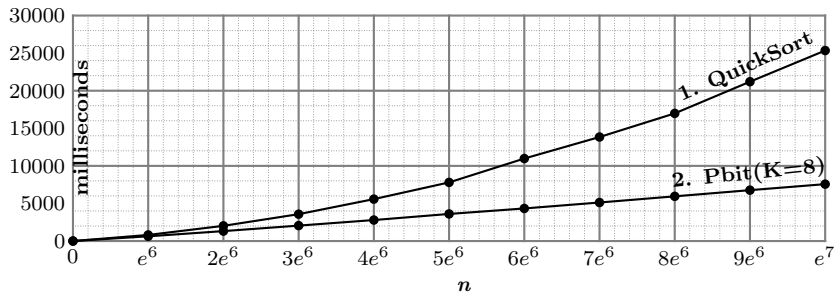
\begin{center}
\begin{footnotesize}
\begin{pspicture*}(-1,-1)(11,3.2)
\psgrid[gridcolor=gray, subgriddots=6](-1,-0.6)(0,0)(10,3)

\rput[b]{24}(9, 2.2){\textbf{1. QuickSort}}
\psline[linearc=0.25, showpoints=true](0,0)(1, 0.0811)(2, 0.2023)(3, 0.3575)(4, 0.5578)(5, 0.7801)(6, 1.0976)(7, 1.3840)(8, 1.6984)(9, 2.1200)(10, 2.5346)

\rput[b]{6}(9, 0.75){\textbf{2. Pbit(K=8)}}
\psline[linearc=0.25, showpoints=true](0,0)(1, 0.0631)(2, 0.1322)(3, 0.2053)(4, 0.2794)(5, 0.3605)(6, 0.4336)(7, 0.5127)(8, 0.5948)(9, 0.6770)(10, 0.7571)

\rput[l](5,-0.6){ \textbf{\emph{n}}}
\rput[c](0,-0.2){$0$}
\rput[c](1,-0.2){$e^6$}
\rput[c](2,-0.2){$2e^6$}
\rput[c](3,-0.2){$3e^6$}
\rput[c](4,-0.2){$4e^6$}
\rput[c](5,-0.2){$5e^6$}
\rput[c](6,-0.2){$6e^6$}
\rput[c](7,-0.2){$7e^6$}
\rput[c](8,-0.2){$8e^6$}
\rput[c](9,-0.2){$9e^6$}
\rput[c](10,-0.2){$e^7$}

\rput[r](-0.1, 0){$0$}
\rput[r](-0.1, 0.5){$5000$}
\rput[r](-0.1, 1){$10000$}
\rput[r](-0.1, 1.5){$15000$}
\rput[r](-0.1, 2){$20000$}
\rput[r](-0.1, 2.5){$25000$}
\rput[r](-0.1, 3){$30000$}

\rput[t]{90}(0, 1.5){\textbf{milliseconds}}
\end{pspicture*}

\begin{tabular}{|c|c|c|c|c|c|c|c|c|c|c|}
\hline
\textbf{\emph{n}} & $e^{6}$ &	$2e^{6}$	&$3e^{6}$&	$4e^{6}$&	$5e^{6}$&	$6e^{6}$&	$7e^{6}$&	$8e^{6}$&	$9e^{6}$	&$e^{7}$\\
\hline
1. &811	&2023	&3575	&5578	&7801	&10976	&13840	&16984	&21200	&25346\\
\hline
2. &631	&1322	&2053	&2794	&3605	&4336	&5127	&5948	&6770	&7571\\
\hline
\end{tabular}
\break For this test I used algorithm quicksort from standard library C\\
\end{footnotesize}
\end{center}

\section{Double-linked lists sorting}
\emph{We use double-linked lists very often. How to adopt Pbit for sorting such lists? }
In case of double-linked lists sorting we do not change Pbit's code. We only add repair function ``reverse connection'' of list elements (nodes). After (double-linked) list sorting using Pbit, pointers showing the previous nodes are incorrect. Function which changes ``reverse connection'' should receive root of the list as parameter. Next, in turns going through its elements it should set up pointer `back' in such a way as to show the element right behind.

\section{Floating-point numbers sorting}

Floating-point numbers in digital computer written in the following form\footnote{The problem of binary arrangements of float-point numbers was well described by Daniel W. Lewis~\cite{Daniel} }
\begin{center}
$l = m * 2^c$, $l$ - number, $m$ - mantissa, $c$ - characteristic
\end{center}
In principle, characteristic is more important than mantissa for the process of sorting. 
This is why we can sort the number in relation to mantissa first, and then once more in relation to its characteristic. Such sorting is correct because Pbit is stable. You should remember about conversion, because logical shift operator does not work with floating-point numbers! 
\begin{small}
\begin{center}
\begin{tabular}{|l|}
\hline
Floating-point formats (standard ANSI IEEE 754): \\
--- simple, single precision (SINGLE): $m=23+1, c=8$\\
--- extended, single precision (SINGLE, EXTENDED): $m >= 32, c >= 11$\\
--- simple, double precision (DOUBLE): $m=52+1, c=11$\\
--- extend, double precision (DOUBLE, EXTENDED): $m >= 64, c >= 15$\\
\hline
\end{tabular}
\end{center}
\end{small}

\section{Portability}

\textbf{Byte order}\footnote{The problem of code portability was well described by Kernighan and Pike~\cite{Ker}.}
It's obvious that all modern machines have 8-bit bytes. But in different machines there are different representations of object greater than one byte. For example, integers type 'short' (language C), which typical have two bytes, may be stored in memory in two ways: 
their less significant byte will be at smaller address (less significant byte first, so little-endian order of bytes) or inversely - at greater address (more significant byte first, so big-endian order of bytes) than more significant byte. Although machines in both cases treat memory as sequence of words in the some order, they interpret the byte order within the words differently. This is why it's important to change the lines of code which is responsible for looking through bits by Pbit (the version of algorithm described in documentation was implemented by machine of little-endian type).

\begin{tabular}{|l|}
\hline
\verb@#include<stdio.h>@\\
\verb@int main(){@\\
\verb@unsigned long x=0x11223344UL;@\\
\verb@unsigned int size=sizeof(x);@\\
\verb@unsigned char *p=(unsigned char *)&x;@\\
\verb@while(size--) printf("%x, ",*p++);@\\
\verb@return 0;@\\
\verb@}@\\
\hline
\end{tabular}
\parbox[c]{28ex}{To check how our machine implements the byte order within the word we can use a simple program (standard PC is machine of little-endian type)}
The result of this program in machine with decreasing order of bytes\break (Motorola 68K): 11, 22, 33, 44,\\
In case of machine with ascending order of bytes\break (80x86 Intel Architecture): 44, 33, 22, 11,\\
Machine type PDP-11 (old-fashioned 16-bit machine) will give the result:\break 22, 11, 44, 33,\\
\\
\textbf{Arithmetic or logical shift}
Shift of value with number sign to the right by means of operator `\verb@>>@' can be treated as arithmetic shift (the copy of bit sign is copied during bit shift) or logical (zero will be placed in released bits during the shift). Fortunately, this problem does not concern Pbit, despite the fact that I use shift operator. It happens that way because operation of logical `\verb@AND@' comes after the operation of bit shift, which unifies the manner of bit shift.
\\
\\
\\
\parbox{\textwidth}{
\textbf{\large{Gratefulness}}
\newline
\emph{
I do extend thanks to Marek Gajecki -- scientist who has been teaching me everything about algorithms.
I want to special thanks to Tomasz Jurczyk for effective implementation QuickerSort and MergeSort.
}
}


%
%
%
\end{document}